# Quench dynamics and quantum flutter properties of one-dimensional attractive single-spin flipped Fermi gases*

YIN Xiangguo[1, 2, 3], ZHAO Shulong[1, 2]

1. Institute of Theoretical Physics, Shanxi University, Taiyuan 030006, China
2. State Key Laboratory of Quantum Optics Technologies and Devices, Shanxi University, Taiyuan 030006, China
3. Collaborative Innovation Center of Extreme Optics, Shanxi University, Taiyuan 030006, China

**Abstract** The nonequilibrium dynamics of impurities in quantum many-body systems constitute a frontier topic in ultracold-atom physics and are key to uncovering the microscopic mechanisms of polaron physics and collective excitation behavior. Compared with the extensively studied repulsively interacting systems, the dynamical behavior under attractive interactions remains insufficiently explored in a systematic manner, owing to the complex coupling mechanisms among bound states, gas states, and irregular states. We take the ground state of a one-dimensional ideal Fermi gas as the initial state, and add a spin-down impurity particle carrying momentum. The impurity has an attractive interaction with the spin-up fermions. We study the time-evolution behavior of the two-body correlation functions between particles of different spins and the impurity momentum. First, based on the exact solution from the Bethe ansatz (BA), starting from the eigenstate wave functions, we analytically simplify the overlap integrals, the matrix elements of the one-body correlation function of spin-up particles, and the matrix elements of the two-body correlation functions between particles of different spins into finite-sum expressions of simple functions, thereby enabling efficient computations of the occupation probabilities of a large number of eigenstates and the long-time evolution of the correlation functions. Second, in the weakly attractive regime: when the total momentum is less than or equal to the Fermi momentum, the dynamics exhibits features of bound states. The two-body correlation function between particles of different spins shows a stable correlation peak and an outward-propagating correlation hole, and its evolution period agrees with the period corresponding to the energy difference between the two bound states with the highest occupation probabilities. Conversely, when the total momentum is greater than the Fermi momentum, due to the superposition of bound states, gas states, and irregular states, the system exhibits complex oscillatory behavior, alternating between bound-state-like and gas-state-like features. In the strongly attractive regime, regardless of the initial momentum, the system exhibits bound-state-dominated features locally, characterized by a localized correlation peak and Friedel-like oscillations induced by scattering interference. Finally, we rigorously characterize the phenomenon of quantum flutter, namely the periodic oscillation of the impurity momentum, and find that the quantum-flutter period is strictly determined by the chemical potential of the spin-down impurity particle $\mu_{\downarrow c}$, namely $\tau = 2\pi/|\mu_{\downarrow c}|$. We obtain consistent results for the chemical potential of the spin-down particle via four approaches: For zero total momentum, it can be obtained by deriving the critical magnetic field and the critical chemical potential at the quantum phase-transition point from the thermodynamic BA equations, and it can also be derived from the discrete BA equations via the energy required to inject a single spin-down particle into the ground state of the ideal Fermi gas; when the total momentum equals the Fermi momentum, the energy difference between the gas-state magnon and the bound-state exciton equals this chemical potential; when the total momentum is greater than the Fermi momentum, by using the relation between the single-particle excitation energy and the dressed energy, we can derive that the energy difference between the bound-state exciton states and the gas-state magnon states, which have relatively high occupation probabilities and form pairwise correspondences, equals this chemical potential. This work investigates the evolution laws of quench dynamics and quantum flutter in systems with attractive interactions, which helps deepen the understanding of nonequilibrium-state properties in quantum many-body physics and provides theoretical support for related ultracold-atom experiments on impurity dynamics.

# 1 Introduction

In recent years, the rapid development of cold atom and quantum simulation experiments has provided new controllable platforms for studying the non-equilibrium dynamics of many-body systems[1-3]. These experimental platforms feature highly tunable interaction strengths and dispersion relations, offering ideal conditions for precise investigations of non-equilibrium dynamics[4,5]. The model describing the interaction between a one-dimensional background particle gas in a specific state and impurity particles has become a research hotspot in recent years[6-9]. Studying moving impurities helps elucidate the rich collective and interference phenomena exhibited in these impurity-containing quantum many-body systems. These phenomena include polarons[2,10,11], Bogoliubov-Cherenkov radiation[12], shock waves[13,14], dissipative quantum impurity dynamics[15], excitation spectra and effective mass[16-18], orthogonality catastrophe[19,20], logarithmic diffusion of Green's functions[21], persistence of threshold singularities in spectral functions[22-24], Bloch oscillations[25,26], quantum flutter[7,27], and anyonization of one-dimensional bosons[28].

As a typical representative of one-dimensional quantum integrable systems, the one-dimensional Bose gas with contact interactions (the Lieb-Liniger model)[7,29] exhibits a phenomenon known as quantum flutter. This phenomenon occurs when the velocity of an impurity injected into the Bose gas exceeds the intrinsic sound velocity. Research indicates that the impurity transfers only part of its momentum to the background gas. Subsequently, it forms a correlated state characterized by long-time momentum oscillations without decay. Quantum flutter was discovered by studying the complete quantum mechanical evolution of polarons. This evolution can be described by exact Bethe ansatz (BA) solutions. The BA method not only provides exact solutions for the Lieb-Liniger model but also finds wide application in other quantum integrable systems[30,31].

Similar to the aforementioned Lieb-Liniger model, another classic model of quantum integrable systems is the one-dimensional Fermi gas with contact interactions (the Yang-Gaudin model), which is the focus of this study. Research on the equilibrium and dynamic properties of Fermi polarons includes spin-charge separation[1], the transition from polaron to molecular states under attractive interactions[32], the transition and coexistence of fermionic polarons and molecular states[33,34], the emergence of sharp quantum collective modes[35], anomalous momentum distributions of impurities[36,37], dynamic properties of the system after shock[38], quantum flutter induced by supersonic impurities[27], and quantum quench dynamics[39]. Currently, research on quantum flutter mainly focuses on systems with repulsive interactions. There is a lack of systematic exploration of quench dynamics and related quantum flutter phenomena under attractive interactions. This gap stems primarily from two challenges. Regarding physical mechanisms, attractive interactions may cause impurities and background particles to form bound pairs. Consequently, non-equilibrium dynamics are influenced jointly by bound states, gas states, and even unconventional states. Their behavioral mechanisms may differ from those in repulsive systems. Regarding computation, the coexistence of multiple types of eigenstates (bound states, gas states, and unconventional states) significantly increases the complexity of analytically deriving and numerically calculating correlation functions and initial state overlap integrals between different eigenstates.

To overcome these difficulties, this study proposes an efficient calculation method based on the exact solution of the Yang-Gaudin model. This method converts matrix elements of correlation functions and initial state overlap integrals into finite sum forms, significantly reducing the computational burden. Using this approach, we conducted a systematic quantitative study of many-body quench dynamics under attractive interactions. We discovered quantum flutter behavior in systems with attractive interactions and explored its physical mechanism. This provides a new theoretical perspective for understanding the role of interaction strength in quantum non-equilibrium evolution. The main contents of this paper are as follows: Section 2 introduces the exact BA solutions and corresponding wave functions for bound and gas states in a single-spin-flip Fermi gas. It further provides the wave functions for unconventional eigenstates. Section 3 presents the expressions for time-evolving wave functions and correlation functions. It simplifies the correlation functions of different eigenstates and initial state overlap integrals into finite sum forms. Section 4 obtains the quench dynamic properties of two-body correlation functions between different spins under various momenta and interaction strengths through numerical calculations. Combining these with the occupation probabilities of a large number of system eigenstates, we analyze the relationship between the period of correlation function evolution after quenching and the energy difference of high-probability states. Section 5 discusses quantum flutter and quantum revival phenomena under different interactions, starting from the time evolution of momentum. It also examines the correspondence between the energy difference associated with the quantum flutter period and the chemical potential of spin-down particles. Finally, Section 6 summarizes the research findings.

# 2 Exact Solution of the Single-Spin-Flip Fermi Gas Model

The Hamiltonian of the single-spin-flip Fermi gas system can be written in the following form:

$$\hat{H} = -\frac{\hbar^2}{2m}\sum_{i=1}^{N}\frac{\partial^2}{\partial x_i^2} + g\sum_{i<j}\delta(x_i - x_j). \tag{1}$$

The system consists of $N-1$ spin-up fermions and $1$ spin-down impurity fermion. The variable $x_i$ denotes the coordinate of the $i$-th fermion, $\hbar$ represents the reduced Planck constant, and $m$ is the mass of a single fermion. The parameter $g$ indicates the interaction strength between fermions with different spins. A positive $g$ corresponds to repulsive interactions, whereas a negative $g$ signifies attractive interactions. This study primarily investigates the properties of the system under attractive interactions. For notational convenience, we set $\hbar = 2m = 1$, yielding the following Hamiltonian for the system:

$$\hat{H} = -\sum_{i=1}^{N} \frac{\partial^2}{\partial x_i^2} + 2c \sum_{i<j} \delta(x_i - x_j), \tag{2}$$

where $c = g/2$.

The Bethe ansatz method solves the eigenvalue equation for this model. Under periodic boundary conditions, let $x_i$ denote the coordinates of spin-down particles. The wave function for the standard eigenstates of the system can be expressed as

$$\begin{aligned} &\psi_{\downarrow i}(x_1, x_2, \cdots, x_N; v_\mathrm{r}) \\ &= \frac{1}{\sqrt{G_{v_\mathrm{r}}}} \sum_P (-1)^P \left[1 - \exp(\mathrm{i}k_{P_i}L)\right] \exp(\mathrm{i}k_{P_i}x_i) \\ &\times \prod_{j=1(j\neq i)}^{N} \exp\left[\mathrm{i}k_{P_j}\left(L\theta(x_i - x_j) + x_j\right)\right], \end{aligned} \tag{3}$$

Here, $v_\mathrm{r}$ denotes the conventional eigenstate indicator; $L$ represents the system length; $\theta(x)$ is the step function, defined as $\theta(x) = 0$ for $x < 0$ and $\theta(x) = 1$ for $x > 0$; $P$ signifies a standard permutation, with $(-1)^P$ being the sign factor arising from the antisymmetry of fermion exchange. When $P = \{1, 2, \cdots, N\}$, $(-1)^P = 1$; swapping any two positions once changes $(-1)^P$ to its opposite. The normalization coefficient for the conventional eigenstate wave function is

$$G_{v_\mathrm{r}} = \int_0^L \mathrm{d}x_1 \cdots \int_0^L \mathrm{d}x_N |\psi_{\downarrow 1}(x_1, x_2, \cdots, x_N; v_\mathrm{r})|^2, \tag{4}$$

Its analytical simplified form has been provided in Reference [39].

Conventional eigenstates include scattering states and bound states. Their wave vectors $k_j$ and rapidities $\lambda$ both satisfy the Bethe Ansatz equations in the following form:

$$\frac{k_j - \lambda + \mathrm{i}c'}{k_j - \lambda - \mathrm{i}c'} = \exp(\mathrm{i}k_j L), \tag{5}$$

$$\prod_{j=1}^{N} \frac{k_j - \lambda + \mathrm{i}c'}{k_j - \lambda - \mathrm{i}c'} = 1, \tag{6}$$

Here, $j = 1,2,\cdots,N$ and $c' = c/2$. The total momentum and energy eigenvalues of the system are $Q = \sum_{j=1}^{N} k_j = 2\pi n_Q/L$ and $E = \sum_{j=1}^{N} k_j^2$, respectively, where $n_Q$ denotes the total momentum quantum number.

All wave vectors $k_j$ in the gaseous state are real numbers. By taking the logarithm of Eqs. (5) and (6) and simplifying the expressions, we obtain a set of Bethe ansatz equations for the gaseous state:

$$k_j L - 2\pi n_j + 2\arctan\frac{k_j-\lambda}{c'} = 0,\quad j = 1,2,\cdots,N, \tag{7}$$

$$\sum_{j=1}^{N}\arctan\frac{k_j-\lambda}{c'} + \pi n_\lambda = 0, \tag{8}$$

Here, $n_j$ and $n_\lambda$ are quantum numbers describing the gaseous state.

A pair of string solutions $k_j$ exists for the bound state. Their real parts are equal, and their imaginary parts are opposite in sign. Let $p$ denote the real part of the string solution, and let $\beta$ denote its imaginary part. The remaining $k_j$ values remain real. Substituting the string solutions into Eqs. (5) and (6), taking the logarithm, and simplifying yields a set of Bethe ansatz equations for the bound state:

$$k_{j'} L + 2\arctan\frac{k_{j'}-\lambda}{c'} - 2\pi n_{j'} = 0,\quad j' = 1,2,\cdots,N-2, \tag{9}$$

$$(p-\lambda)^2 + c'^2 + \beta^2 + \frac{2\beta c'}{\tanh(\beta L)} = 0, \tag{10}$$

$$\tan(pL)\cdot(\beta^2 - c'^2 + (p-\lambda)^2) - (p-\lambda)2c' = 0, \tag{11}$$

$$2p + \sum_{j'=1}^{N-2} k_{j'} - \frac{2\pi n_{\lambda'}}{L} = 0, \tag{12}$$

Here, $n_{j'}$ and $n_{\lambda'}$ are quantum numbers describing bound states. Solving these two sets of equations yields the wave vectors and rapidities for all conventional eigenstates of the system.

When the total momentum of the system exceeds the Fermi momentum,

unconventional eigenstates emerge. The wave vectors of these states remain independent of the interaction strength. When the quantum number $n_\lambda$ takes the value $N/2$, the wave vectors are obtained by solving the Bethe Ansatz equations (7) and (8). For $j = 1, \cdots, N-1$, $k_j = 2\pi n_j / L$, where $n_j = -N/2+1, \cdots, N/2-1$. For $j = N$, $k_N = Q$. The wave function of this state resembles that of free fermions, as detailed below:

$$\psi_{\downarrow i}(x_1, x_2, \cdots, x_N; v_{\text{ir}}) = \frac{\text{i}}{N!\sqrt{NL^N}} \sum_P (-1)^P \exp\left(\text{i} \sum_{j=1}^{N} k_{P_j} x_j\right), \tag{13}$$

Here, $v_{\text{ir}}$ denotes the unconventional eigenstate indicator.

## 3 Analytical Calculation of Quench Dynamics and Correlation Functions for Different Eigenstates

The initial wave function of the system is obtained by multiplying the Slater determinant of the $N-1$ spin-up fermion system by the wave function of a single spin-down particle:

$$\psi_{\text{tot}}(0) = \psi^0_{\downarrow 1}(x_1, x_2, \cdots, x_N) = \frac{1}{N! L^{\frac{N}{2}}} \sum_P (-1)^P \exp\left(\text{i} \sum_{j=1}^{N-1} q_{P_j} x_{j+1}\right) \exp\left(\text{i} Q x_{\downarrow 1}\right), \tag{14}$$

The wave vector of the spin-down particle is $Q$, with the corresponding wave function given by $\exp(\text{i}Qx_{\downarrow 1})$. The wave vectors of the spin-up particles are $q_j = (j - N/2) \times 2\pi/L$, where $j = 1, 2, \cdots, N-1$, thereby forming the distribution of the ground-state Fermi sea.

At $t = 0$, the interaction strength between particles of different spins is quenched from $0$ to a finite value $c$. The subsequent time evolution of the system is primarily determined by the interaction strength $c$ and the initial momentum $Q$ of the impurity particle. To investigate the dynamical properties of the system, we apply the time-evolution operator to the initial state wave function to obtain the time-dependent wave function:

$$\psi_{\text{tot}}(t) = \exp(-\mathrm{i}\hat{H}t)\psi_{\text{tot}}(0), \qquad (15)$$

Inserting the complete basis of the system after the time-evolution operator yields:

$$\psi_{\text{tot}}(t) = \sum_{v} C_v \exp(-\mathrm{i}E_v t)\,\psi_v, \qquad (16)$$

Here, $v$ denotes the indices of all eigenstates of the system, and $E_v$ represents the corresponding energy eigenvalues. The overlap integral is defined as

$$\begin{aligned} C_v = \ & \int_0^L \mathrm{d}x_1 \cdots \int_0^L \mathrm{d}x_N \psi^*_{\downarrow 1}(x_1, x_2, \cdots, x_N; v) \\ & \times \psi^0_{\downarrow 1}(x_1, x_2, \cdots, x_N), \end{aligned} \qquad (17)$$

The square of its modulus represents the occupation probability. In numerical calculations, eigenstates with low occupation probabilities are neglected, satisfying the condition $\sum_v |C_v|^2 > 0.95$.

Substituting the wave function of an arbitrary conventional eigenstate into Eq. (17) and simplifying using momentum conservation and the BA equations reduces the overlap integral to a product of simple functions:

$$\begin{aligned} C_{v_{\mathrm{r}}} = & (-1)^{\xi} \frac{-\mathrm{i}N!\,(-2)^N}{\sqrt{L^{N-2}N!NG_{v_{\mathrm{r}}}}} \prod_{j=1}^{N} \sin(k_j L/2) \\ & \times \exp(-\mathrm{i}QL/2)\det\left(M^{\mathrm{C}}\right). \end{aligned} \qquad (18)$$

When $v_{\mathrm{r}}$ in the equation belongs to a bound state (gaseous state), $(-1)^{\xi} = -1(1)$. Here, $k_j$ represents the wave vectors of different particles in an eigenstate, and $\boldsymbol{M}^{\mathbf{C}}$ is an $N \times N$ matrix. For $i = 1, 2, \cdots, N$ and $j = 1, \cdots, N-1$, the matrix elements are given by $M^{\mathrm{C}}_{ij} = 1/(q_j - k_i)$. All matrix elements in the $N$th column equal 1, and $det()$ denotes the determinant of the corresponding matrix.

Substituting the unconventional eigenstate wave function from Eq. (13) into Eq. (17) to calculate the overlap integral between this state and the initial state yields a constant occupation probability of $1/N$. The occupation probability of the unconventional eigenstate remains independent of the interaction strength. Such states appear not only in the Yang-Gaudin model but also have analogous eigenstates in the Lieb-Wu model[23].

Using the time-dependent wave function, we can express the expectation values of the single-particle correlation function for spin-up particles and the two-particle correlation function for particles with different spins as follows

$$
\begin{aligned}
g_{\uparrow}^{(1)}(x, x'; v, v'; t) = & \sum_{vv'} C_v^* C_{v'} \exp\left[\mathrm{i}\left(E_v - E_{v'}\right) t\right] \\
& \times g_{\uparrow}^{(1)}(x, x'; v, v'), \qquad (19)
\end{aligned}
$$

$$
\begin{aligned}
g_{\uparrow\downarrow}^{(2)}(x, x'; v, v'; t) = & \sum_{vv'} C_v^* C_{v'} \exp\left[\mathrm{i}\left(E_v - E_{v'}\right) t\right] \\
& \times g_{\uparrow\downarrow}^{(2)}(x, x'; v, v'), \qquad (20)
\end{aligned}
$$

The time-independent one-body and two-body correlation functions can be expressed in terms of the wave function as follows

$$
\begin{aligned}
& g_{\uparrow}^{(1)}(x, x'; v, v') \\
= & (N-1) \int_0^L \mathrm{d}x_1 \cdots \int_0^L \mathrm{d}x_{N-1} \psi_{\downarrow 1}^*(x_1, \cdots, \\
& x_{N-1}, x; v) \times \psi_{\downarrow 1}(x_1, \cdots, x_{N-1}, x'; v'), \qquad (21)
\end{aligned}
$$

$$
\begin{aligned}
& g_{\uparrow\downarrow}^{(2)}(x, x'; v, v') \\
= & (N-1) \int_0^L \mathrm{d}x_3 \cdots \int_0^L \mathrm{d}x_N \psi_{\downarrow 1}^*(x, x', x_3, \cdots, \\
& x_N; v) \times \psi_{\downarrow 1}(x, x', x_3, \cdots, x_N; v'). \qquad (22)
\end{aligned}
$$

Therefore, by evaluating Eqs. (21) and (22), one can obtain the analytical expressions for the single-particle correlation function of spin-up particles and the two-particle correlation function between particles with different spins, respectively.

The analytical result for the single-particle correlation function of spin-up particles between any two conventional eigenstates can be simplified to

$$
\begin{aligned}
g_{\uparrow}^{(1)}(x, x'; v_{\mathrm{r}}, v'_{\mathrm{r}}) = (-1)^{\xi} \frac{4(N-1)!}{\sqrt{G_{v_{\mathrm{r}}} G_{v'_{\mathrm{r}}}}} \sum_{m', s=1}^{N} A_{m'}^{s} \\
\times F_a\left(k_{m'}, k'_s, z\right), \qquad (23)
\end{aligned}
$$

where

$$
\begin{aligned}
A_{m'}^{s} = (-1)^{m'+s} \sum_{l=1\ (l \neq m')}^{N} (-1)^{N+l} \operatorname{sgn}(l - m') \\
\times \sin\left(k_l L/2\right) \det\left(\boldsymbol{B}_{m'l}^{s}\right). \qquad (24)
\end{aligned}
$$

When $v_r$ in Eq. (23) belongs to a bound state (gaseous state), $(-1)^{\xi}=-1(1)$. Let $z=x-x', x'=0$, where $k$ and $k'$ in Eq. (23) are the wave vectors of any two regular eigenstates, respectively. During numerical calculations, $k$ and $k'$ must each satisfy a fixed ordering of the regular eigenstate wave vectors. For instance, all real solutions are arranged in ascending order, followed by the two string solutions, which are ordered by their imaginary parts in ascending order. $F_a(k_{m'}, k'_s, z)$ can be specifically expressed as

$$F_a(k,k',z) = \frac{2}{k-k'}\{\exp[ik'(z-L/2)]\sin(kL/2) \\ -\exp[ik(z-L/2)]\sin(k'L/2)\}. \tag{25}$$

Define the $(N+1)\times N$ matrix $\boldsymbol{M}$, where the matrix elements in the first $N$ rows are given by $M_{ij}=\sin[(k'_j-k_i)L/2]/[(k'_j-k_i)/2]$, and the elements in the last row are $M_{N+1,j}=\sin(k'_jL/2)$. The term $\boldsymbol{B}^s_{m'l}$ denotes an $(N-1)\times(N-1)$ square matrix. This matrix is formed by the remaining elements of $\boldsymbol{M}$ after removing the $m'$-th and $l$-th rows as well as the $s$-th column, while preserving their original positions.

The single-particle correlation function for spin-up particles between conventional and unconventional eigenstates can be simplified as

$$g^{(1)}_{\uparrow}(x,x';v_{\mathrm{r}},v'_{\mathrm{ir}}) = \mathrm{i}(-1)^{\xi}\frac{(N-1)!}{\sqrt{G_{v_{\mathrm{r}}}N!L^N}} \\ \times\sum_{m',s=1}^{N}\exp(\mathrm{i}k'_s z)\,F_b(k'_s,k_{m'},L)\,D^s_{m'}, \tag{26}$$

where

$$D^s_{m'} = (-1)^{m'+s}\sum_{l=1\ (l\neq m')}^{N}(-1)^{l+N}\operatorname{sgn}(l-m') \\ \times[1-\exp(-\mathrm{i}k_lL)]\det(\boldsymbol{R}^s_{m'l}). \tag{27}$$

Define an $(N+1)\times N$ matrix $\boldsymbol{S}$, where the matrix elements in the first $N$ rows are

$$S_{ij}=\sin[(k'_j-k_i)L/2]\exp\frac{\mathrm{i}(k'_j-k_i)L/2}{(k'_j-k_i)/2}, \tag{28}$$

The matrix element in the last row is $S_{N+1,j}=1$. $\boldsymbol{R}^s_{m'l}$ is an $(N-1)\times(N-1)$ square matrix. It is formed by the remaining matrix elements of the $\boldsymbol{S}$ matrix after removing the $m$' and $l$ rows and the $s$ column, arranged according to their original positions. $k$ and $k'$ must satisfy the fixed ordering of wave vectors for a set of regular eigenstates and a set of unconventional eigenstates, respectively. The specific form of $F_b(k', k, L)$ is identical to that in Eq. (28).

The single-particle correlation function for spin-up particles between unconventional eigenstates can be simplified as

$$g_{\uparrow}^{(1)}(x, x'; v_{\mathrm{ir}}, v_{\mathrm{ir}}) = \frac{N-1}{NL} \sum_{l}^{N} \exp\left(-\mathrm{i} k_l z\right). \quad (29)$$

By combining Eqs. (23), (26), and (29), one can calculate the single-particle correlation function for spin-up particles in any eigenstate.

The analytical results for the two-particle correlation function between particles with different spins in any two conventional eigenstates can be simplified to

$$g_{\uparrow\downarrow}^{(2)}(x, x'; v_{\mathrm{r}}, v_{\mathrm{r}}) = (-1)^{\xi} \frac{4(N-1)!}{\sqrt{G_{v_{\mathrm{r}}} G_{v_{\mathrm{r}}}}} \sum_{m', s=1}^{N} A_{m'}^{s}$$

$$\times \exp\left[\mathrm{i}\left(k_s' - k_{m'}\right)\left(L\theta(z) - z - \frac{L}{2}\right)\right]. \quad (30)$$

The two-body correlation functions for particles with different spins in conventional and unconventional eigenstates can be simplified as

$$g_{\uparrow\downarrow}^{(2)}(x, x'; v_{\mathrm{r}}, v_{\mathrm{ir}}) = \mathrm{i}(-1)^{\xi} \frac{4(N-1)!}{\sqrt{G_{v_{\mathrm{r}}} N! L^{N}}}$$

$$\times \sum_{m', s=1}^{N} \exp\left[\mathrm{i}\left(k_s' - k_{m'}\right) z\right] V_{m'}^{s}, \quad (31)$$

$$V_{m'}^{s} = 2\mathrm{i}(-1)^{m'+s} \sum_{l(l \neq m')} (-1)^{l} \operatorname{sgn}(l - m')$$

$$\times \sin\left(\frac{k_l L}{2}\right) \exp\left(\frac{-\mathrm{i} k_l L}{2}\right) \det(\boldsymbol{W}_{m'l}^{s}). \quad (32)$$

Define an $(N+1) \times N$ matrix $\boldsymbol{U}$, where the matrix elements in the first row are $U_{1,j} = 1$. The matrix elements $U_{ij}$ from the second row to the $N$th row have the same form as in Eq. (28). $\boldsymbol{W}^{s}_{m'l}$ is an $(N-1)\times(N-1)$ square matrix. It is formed by the remaining matrix elements after removing the $m'$th and $l$th rows and the $s$th column from the $\boldsymbol{U}$ matrix, while maintaining their original positions.

The two-body correlation function between particles with different spins in unconventional eigenstates can be simplified as

$$g^{(2)}_{\uparrow\downarrow}(x,x';v_{\mathrm{ir}},v_{\mathrm{ir}}) = \frac{1}{L^2}\left[(N-1)-N\sum_{l_1=1}^{N}\exp\left(-\mathrm{i}k_{l_1}z\right)\right.$$
$$\left.\times\sum_{l_2=1;\, l_2\neq l_1}^{N}\exp\left(\mathrm{i}k_{l_2}z\right)\right]. \qquad (33)$$

By combining Eqs. (30), (31), and (33), one can calculate the two-body correlation functions for particles with different spins in any eigenstate.

# 4 Quench dynamics of correlation functions

Using the analytical expressions for occupation probabilities and correlation functions derived in Section 3, this section investigates the dynamical properties of the system after a quench. Figure 1(a) illustrates the time evolution of the two-body correlation function between different spins under weak interactions, where the impurity momentum equals the Fermi momentum. Although the spin-down particle is in motion, we can fix the coordinate system to this particle by setting its position to $x'=0$. This approach allows us to study the motion of the associated spin-up particle relative to this origin. When the interaction strength between particles of different spins is weak ($c=-1$), the influence of bound states under attractive interactions leads to significant probability density peaks at $x=0$ and $x=L$. These peaks remain greater than $1$ throughout the time evolution, indicating the formation of bound pairs. Meanwhile, due to particle number conservation, a correlation hole appears to the right of the peak, as shown in Figure 1(b). This hole gradually moves to the right and away from its initial position over time. Its motion characteristics reflect, to some extent, the relative motion of the spin-down particle among the spin-up particles. Specifically, at $t=t_{\mathrm{F}}$, the hole appears near $x/L=0.05$ and moves to the right as time evolves. By $t=30t_{\mathrm{F}}$, the hole reaches approximately $x/L=0.6$. Since the system employs periodic boundary conditions, the hole returns to near $x/L=0.05$ at $t=50t_{\mathrm{F}}$. This motion forms the dark blue periodic stripes shown in Figure 1(a).

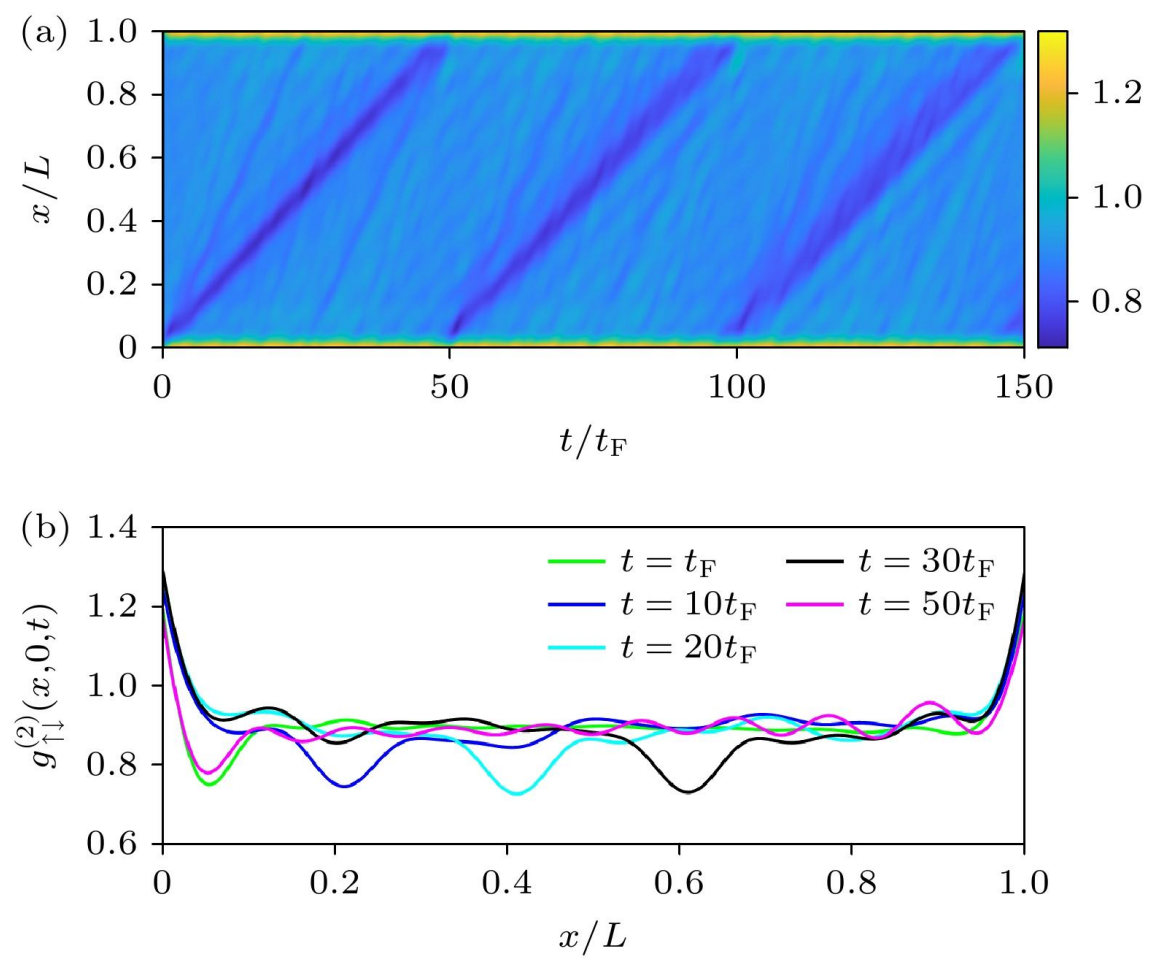

**Fig. 1 The time evolution of the opposite-spin two-body correlation function $g^{(2)}_{\uparrow\downarrow}(x,0,t)$ for total momentum $Q=k_F$. The particle number is $N=10$, the system length is $L=10$, and the interaction strength is $c=-1$. The Fermi momentum is defined as $k_{\mathrm{F}}=(N/2-1)2\pi/L$, the Fermi energy as $E_{\mathrm{F}}=k_{\mathrm{F}}^2$, and the Fermi time as $1/E_{\mathrm{F}}$.**

Figure 2 illustrates the scenario where the momentum of spin-down particles at the initial time exceeds the Fermi momentum. Compared with Figure 1, although the parameters differ only in total momentum, the exhibited properties are distinctly different. In Figure 1, the correlation function at the correlation position $x = 0$ remains greater than $1$ at the initial time. In contrast, the value of the correlation function near $x = 0$ in Figure 2 undergoes drastic changes at the initial time. As time evolves, peaks and troughs alternate at $x = 0$. Peak values greater than $1$ indicate bound-state characteristics, while trough values less than $1$ reflect gaseous-state characteristics. This phenomenon can be understood as follows: the formation of a peak in the correlation function corresponds to the impurity particle capturing a background particle for joint motion. However, due to the high initial kinetic energy of the impurity particle, the bound pair may dissociate during subsequent motion. This dissociation causes the peak to transform into a trough, indicating that the impurity particle temporarily escapes the confinement of the background particle. In Figure 2(b), a distinct correlation hole forms to the right of $x = 0$, and this hole does not propagate outward. A correlation hole also forms to the left of $x = L$, but its formation speed is slower, which differs significantly from the case below the Fermi momentum. A peak forms in the positive direction of the correlation hole, as shown by the irregular yellow stripes with a period of $25t_{\mathrm{F}}$ in Figure 2(a). This peak propagates to the right, consistent with the direction of hole motion in Figure 1(b). Specifically, at $t = t_{\mathrm{F}}$, a small peak appears near $x/L = 0.18$ in Figure 2(b), and subsequently moves to the right. At $t = 16t_{\mathrm{F}}$, the peak reaches approximately $x/L =$ $0.8$. Due to periodic boundary conditions, the peak returns to near $x/L = 0.2$ at $t = 21t_{\mathrm{F}}$. This phenomenon resembles the repulsive case and is associated with a significant increase in the occupation probability of the gaseous state at high momentum. Notably, the correlation hole to the left of $x = L$ exhibits leftward propagation, as indicated by the blue stripes with a period of approximately $50t_{\mathrm{F}}$ in Figure 2(a).

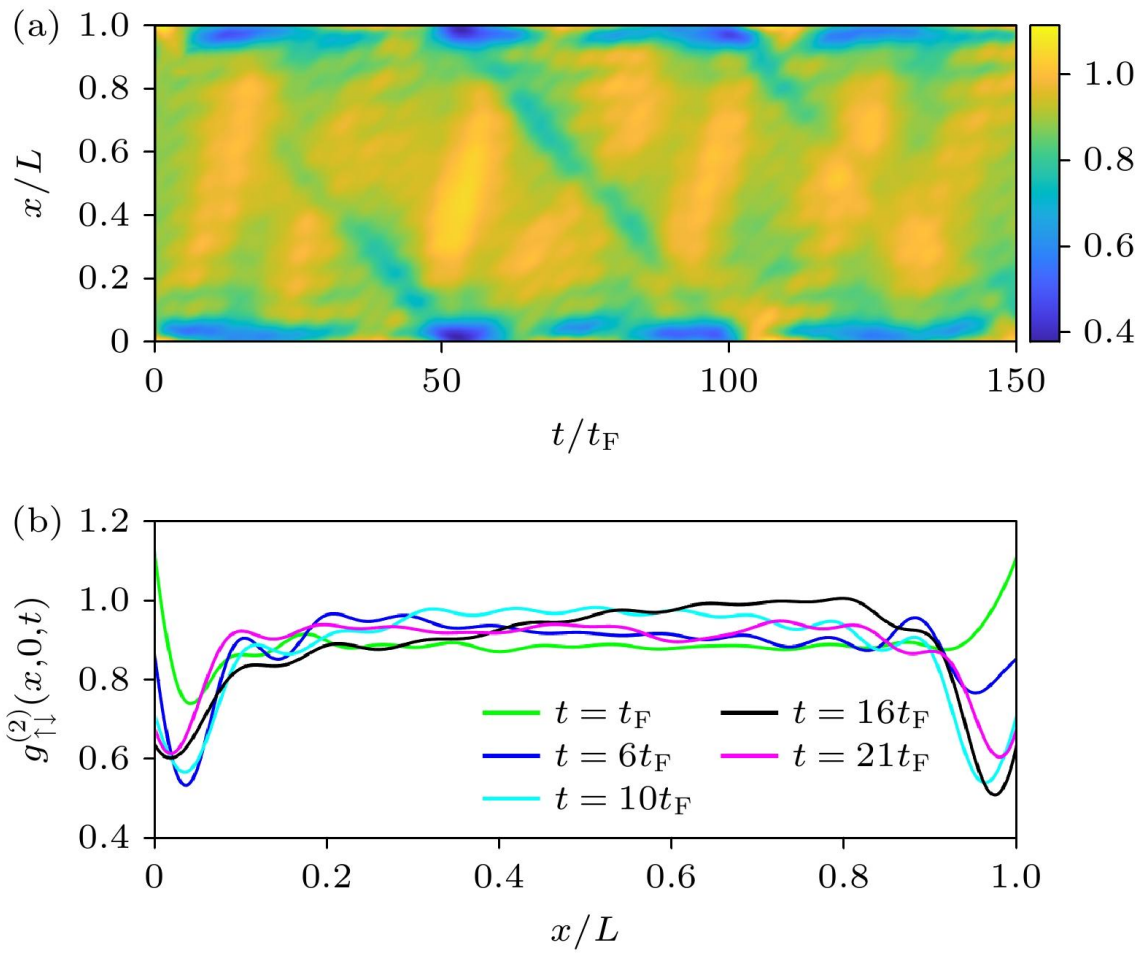

**Fig. 2 The time evolution of the opposite-spin two-body correlation function $g^{(2)}_{\uparrow\downarrow}(x,0,t)$ for total momentum $Q$=1.5$k_{\mathrm{F}}$. The particle number is $N$=10, the system length is $L$=10, and the interaction strength is $c$=−1.**

To elucidate the evolutionary characteristics of the aforementioned correlation functions, Figure 3 presents the distribution of occupation probability as a function of energy for $c = -1$. For a given total momentum $Q$, the energies corresponding to the two eigenstates with the highest occupation probabilities are denoted as $E_{\mathrm{large1}}$ and $E_{\mathrm{large2}}$. The absolute difference between these two eigenenergies is defined as $|\delta E_{\mathrm{large}}|$, and the corresponding characteristic time is $t_{\mathrm{large}} = 2\pi/|\delta E_{\mathrm{large}}|$, which reflects the evolution period of the two-body correlation functions for different spins. When $Q \leqslant k_{\mathrm{F}}$, as shown in Figure 3(a), the two eigenstates with the highest occupation probabilities are bound states. Consequently, the values of the different-spin two-body correlation functions at $x = 0$ in Figure 1(b) remain greater than $1$, primarily exhibiting the properties of bound states. The characteristic time is $t_{\mathrm{large}} = 47.81t_{\mathrm{F}}$, which is basically consistent with the period of the spatiotemporal evolution of the correlation hole in Figure 1. When $Q > k_{\mathrm{F}}$, as illustrated in Figure 3(b), the two eigenstates with the highest occupation probabilities correspond to the gas state and the bound state, respectively, and unconventional eigenstates with relatively high occupation probabilities also exist. Therefore, the values of the different-spin two-body correlation functions at $x = 0$ in Figure 2(b) evolve from greater than $1$ to less than $1$. This indicates that the system dynamics gradually shift from being dominated by bound state contributions to being dominated by gas state contributions. The characteristic time is $t_{\mathrm{large}} = 36.45 t_{\mathrm{F}}$, which no longer aligns with the period of peak evolution in Figure 2. This phenomenon arises because, when the momentum exceeds the Fermi momentum, the weight of the highly occupied bound states decreases significantly. The total occupation probability of bound states becomes lower than that of gas states, while unconventional eigenstates begin to emerge, thereby exhibiting complex coupling characteristics among bound states, gas states, and unconventional eigenstates.

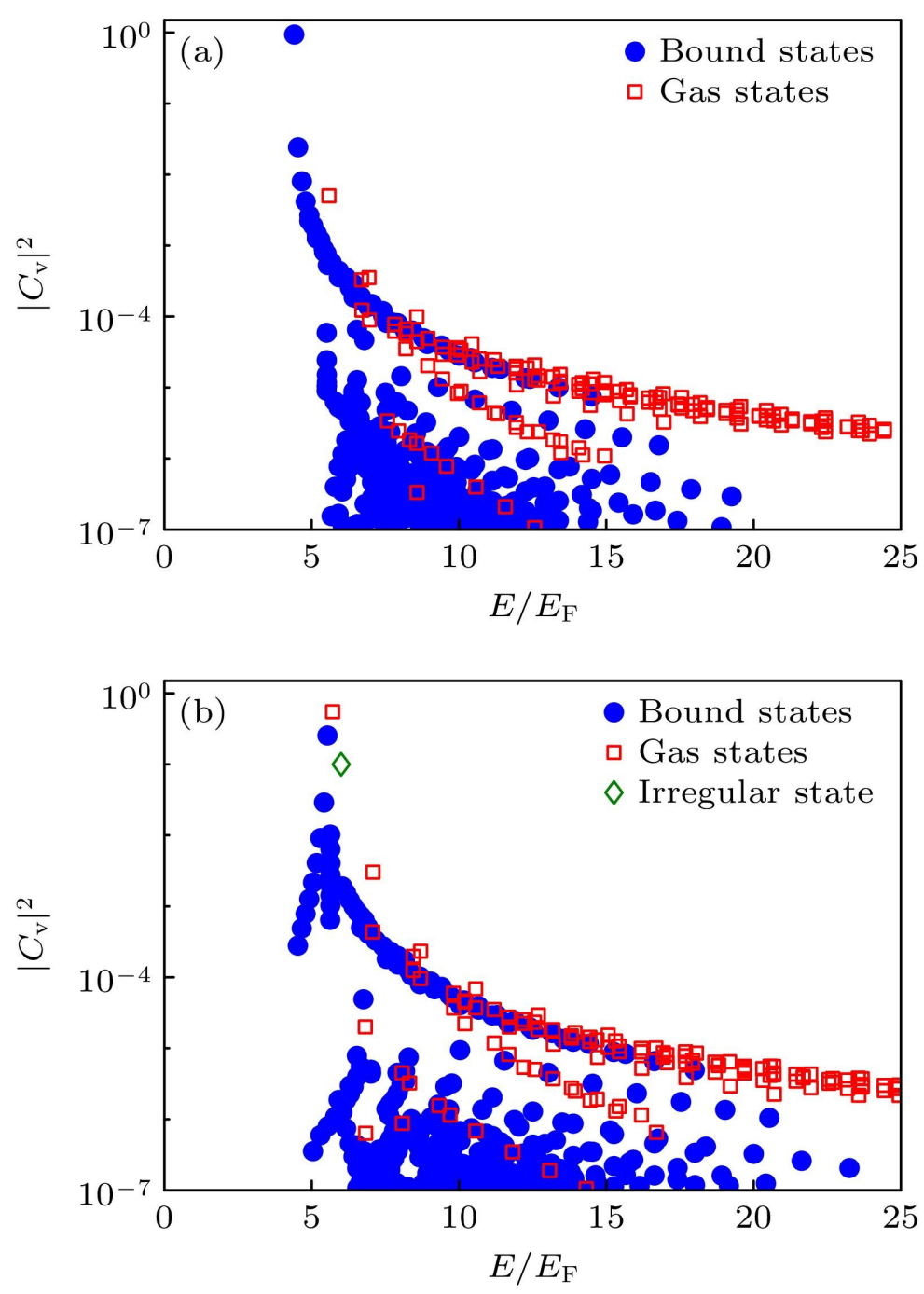

**Fig. 3 The energy distribution of the occupation probabilities of different eigenstates for total momentum (a) $Q=k_{\mathrm{F}}$ and (b) $Q=1.5k_{\mathrm{F}}$. The particle number is $N=10$, the system length is $L=10$, and the interaction strength is $c=-1$.**

When the interaction strength is strong ($c = -10$), as shown in Figure 4, the two-body correlation functions for different spins consistently exhibit a significant peak at the correlation point $x = 0$, regardless of whether the total system momentum is less than or equal to or greater than the Fermi momentum. This feature reflects the localized nature of the bound state. The physical reason is that the local correlation function of the bound state is positively correlated with the absolute value of the interaction strength when the interaction is strong. Consequently, the bound state makes a more dominant contribution throughout the time evolution. As shown in Figure 4(a), correlation holes appear on both sides of the peak at $x = 0$, and these holes do not propagate outward. At $t = t_F$ and $x/L = 0.2$, the peak amplitude in Figure 4(a) is smaller than the corresponding peak in Figure 4(b). This indicates that increased momentum makes the gaseous characteristics of the system more pronounced. The peak in Figure 4(a) at $x/L = 0.2$ propagates to the right with almost no attenuation. In contrast, the peak at the same position in Figure 4(b) propagates to the right, while a smaller trough to the left of $x = L$ propagates to the left. These two waves meet at approximately $4t_F$ and form oscillations. These oscillations resemble the two-body correlations of background particles and are consistent with the phenomena observed in repulsive Fermi gases in Reference [27]. They are referred to as Friedel-like oscillations.

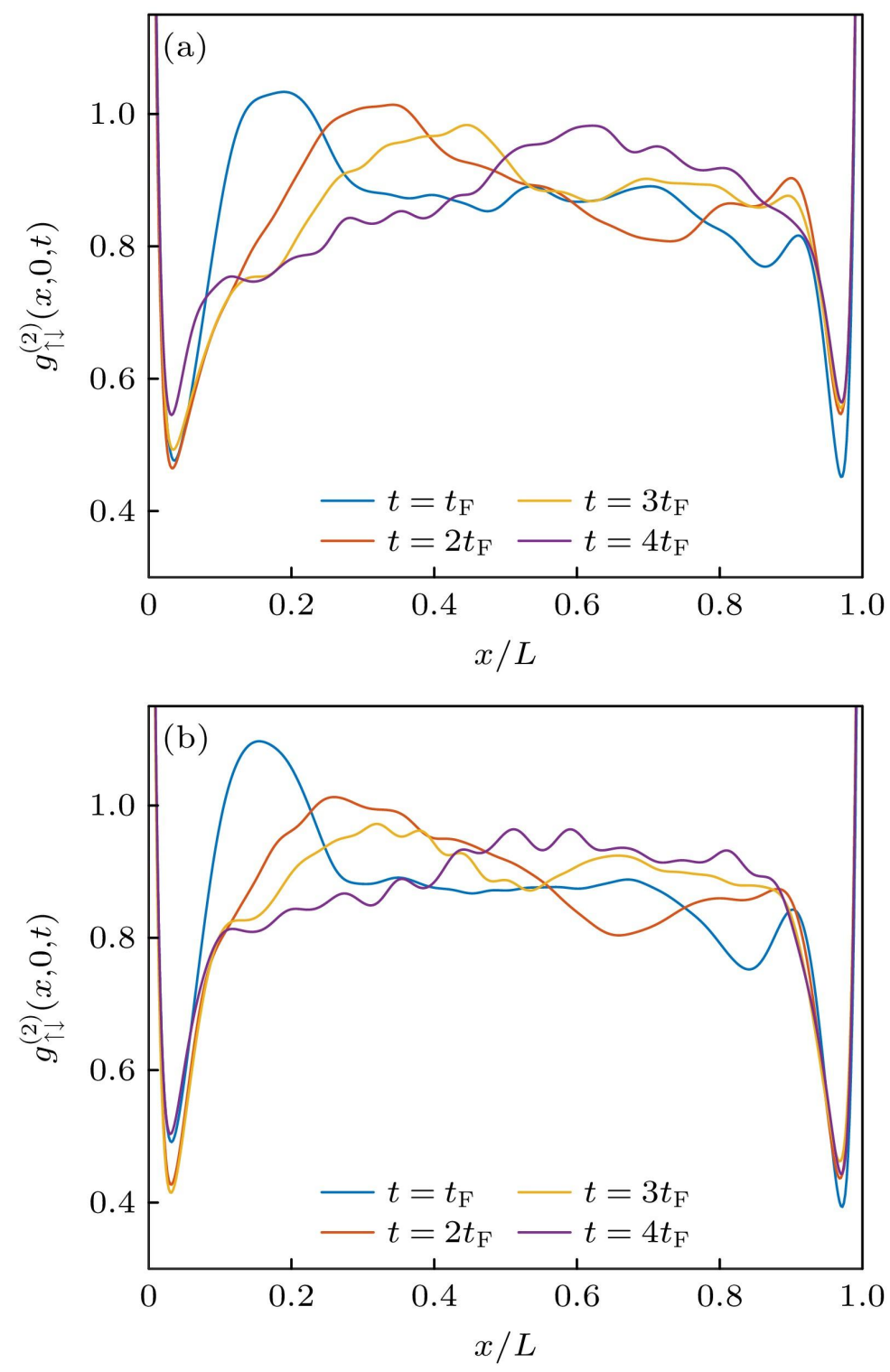

**Fig. 4 The time evolution of the opposite-spin two-body correlation function $g^{(2)}_{\uparrow\downarrow}(x,0,t)$ for total momentum (a) $Q=k_F$ and (b) $Q=1.5k_F$. The particle number is $N=10$, system length is $L=10$, and interaction strength is $c=-10$. The maximum value of the correlation function at $x=0$ is clipped to 1.15.**

# 5 Momentum Evolution and Quantum Flutter of Spin-Down Particles

The momentum distribution of spin-up particles can be obtained by performing a

Fourier transform on the single-particle correlation function of spin-up particles:

$$n_{\uparrow}(k_{m'}) = \frac{L}{2\pi}\int_0^L \mathrm{d}z\, g_{\uparrow}^{(1)}(0,z)\exp(\mathrm{i}k_{m'}z), \tag{34}$$

where $k_{m'} = 2\pi m'/L$, and $m'$ is an integer. Through momentum conservation

$$Q_{\downarrow} = Q - \sum_{k_{m'}} k_{m'} n_{\uparrow}(k_{m'}), \tag{35}$$

We derive the time evolution of the momentum for spin-down particles. Subsequently, we systematically analyze the momentum evolution behavior of spin-down particles and establish the relationship between the period of quantum flutter and the chemical potential of spin-down particles.

## 5.1 Evolution Behavior of Spin-Down Particle Momentum

To facilitate a comparison of momentum evolution differences under repulsive and attractive interactions, Figure 5 presents the time evolution of the momentum for spin-down particles in both scenarios. In the Yang-Gaudin model with repulsive interactions, the evolution exhibits phenomena similar to those in the Lieb-Liniger model[7,29]. As shown in Figure 5(b), under repulsive interactions, the system momentum decays over time to a stable value when the initial total momentum is less than the Fermi momentum (indicated by the blue solid line before $30t_{\mathrm{F}}$). When the total momentum exceeds the Fermi momentum, quantum flutter occurs with a period of approximately $9.7t_{\mathrm{F}}$. In contrast, Figure 5(a) shows that the involvement of bound-state string solutions in the attractive case leads to momentum evolution results that differ from the repulsive case. Regardless of whether the initial total momentum exceeds the Fermi momentum, the time evolution of momentum exhibits quantum-flutter-like phenomena. Specifically, when the initial momentum of the impurity particle exceeds the Fermi momentum, a first secondary peak appears at approximately $t = 3t_{\mathrm{F}}$ (position of dashed line c in Figure 5(a)), and a second secondary peak appears at approximately $t = 6t_{\mathrm{F}}$ (position of dashed line e in Figure 5(a)), irrespective of the specific initial momentum value. The interval between these two secondary peaks corresponds to the period of quantum flutter, which is approximately $3t_{\mathrm{F}}$. The period of quantum flutter is independent of the magnitude of the initial system momentum.

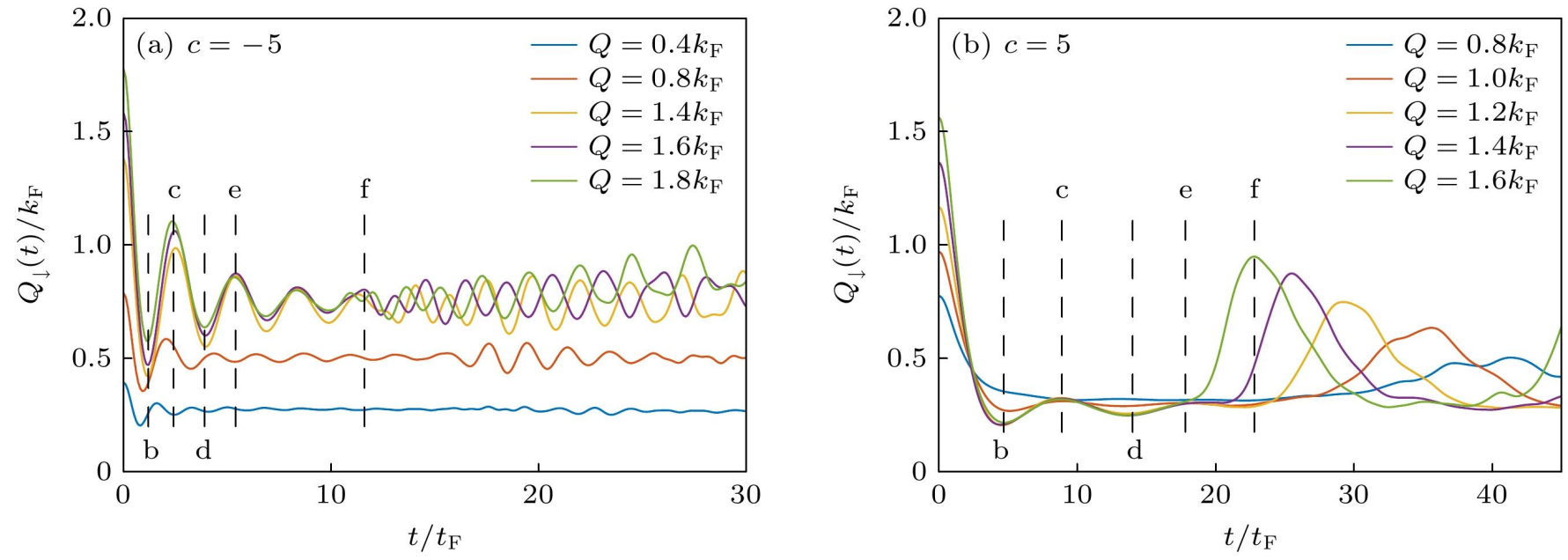

**Fig. 5 Time evolution of the momentum of the spin-down particle for (a) attractive and (b) repulsive interactions. The particle number is $N$=22 and the system length is $L$=22: (a) For interaction strength $c$=−5, the total momentum is $Q$=0.4$k_F$, 0.8$k_F$, 1.4$k_F$, 1.6$k_F$, and 1.8$k_F$; (b) for interaction strength $c$=5, the total momentum is $Q$=0.8$k_F$, $k_F$, 1.2$k_F$, 1.4$k_F$, and 1.6$k_F$. The black dashed lines mark several characteristic times for $Q$=1.6$k_F$: the first time when the impurity momentum reaches a minimum (dashed line b), the first appearance of a secondary peak (dashed line c), the second time when it reaches a minimum (dashed line d), the second appearance of a secondary peak (dashed line e), and the time at which quantum revival occurs (dashed line f).**

To further understand the momentum evolution of spin-down particles over longer time scales, we next investigate quantum revivals under different interaction scenarios. As shown in Fig. 5(b), a main peak appears at $22.81t_F$ for $Q = 1.6k_F$ under repulsive interactions, indicating the occurrence of quantum revival in the system. This phenomenon arises because the system employs periodic boundary conditions, allowing particles to travel one system length at a velocity approximately equal to their initial momentum and return to their initial positions. The time of occurrence can be approximated as

$$t_{\text{revival}} \approx \frac{Lm}{\hbar Q} = \frac{\pi (N/2-1)^2}{n_Q} t_F. \qquad (36)$$

Using Eq. (36), we calculate the quantum revival time for $Q = 1.6k_F$ as $19.63t_F$, which corresponds to the position of the main peak in Fig. 5(b). A larger total momentum shortens the time required for particles in the system to complete a cycle and return to their initial positions. This behavior differs from quantum trembling, which occurs only when the total momentum exceeds the Fermi momentum. In contrast, quantum revival occurs regardless of whether the initial total momentum exceeds the Fermi momentum. In the attractive case shown in Fig. 5(a), the momentum evolution of particles begins to exhibit periodic oscillations after $12t_F$. The amplitude initially increases and then decreases. However, the maximum amplitude is lower than that of the preceding quantum trembling and also lower than the maximum momentum increment observed during quantum revival in the repulsive case. This indicates that the system exhibits phenomena resembling quantum revival due to periodic boundary conditions. Nevertheless, the revival effect is significantly weaker than in the repulsive case with the same interaction strength due to the influence of bound pairs.

The differences in momentum evolution over time for repulsive and attractive interactions are reflected not only in the momentum evolution of spin-down particles but also in the momentum distribution of spin-up particles. Figure 6 presents the momentum distributions of spin-up particles at the six specific moments identified in Fig. 5. These selected moments are: the initial time, the time when the impurity particle momentum first reaches a minimum, the time when the first secondary peak appears, the time when the second minimum is reached, the time when the second secondary peak appears, and the time when quantum revival occurs. These correspond to the initial position and the positions marked by dashed lines b through f in Fig. 5, respectively. For the repulsive interaction system, the red circles in Fig. 6 show that the momentum of spin-up particles is distributed only within the Fermi sea at the initial time. At time $t = 4.7t_F$, the momentum evolution of spin-down particles reaches its first minimum. Due to the repulsive interaction, the momentum distribution of spin-up particles differs significantly from the initial state of the system. A high peak appears at $k_{m'} = 1.8k_F$. This indicates that spin-down particles excite some spin-up particles to high-energy states through repulsive interactions. The peak position is slightly higher than the momentum of the injected spin-down particles at the initial time, $Q = 1.6k_F$. At time $t = 8.9t_F$, the momentum of spin-down particles reaches the first secondary peak, and the momentum distribution of spin-up particles within the Fermi sea near the initial momentum direction contracts significantly. At time $t = 14.0t_F$, the momentum of spin-down particles reaches the second minimum, and the momentum distribution of spin-up particles shows more pronounced contraction near the Fermi sea in the direction of the initial momentum. At time $t = 22.8t_F$, the momentum of spin-down particles exhibits a high peak, indicating that quantum revival occurs and the system approaches its initial state. The momentum distribution of spin-up particles resembles that at the initial time. The peak near $k_{m'} = 1.6k_F$ basically disappears, with significant differences from the initial state remaining only near the Fermi sea. This suggests that spin-up particles have essentially returned their momentum to spin-down particles through interactions at this stage.

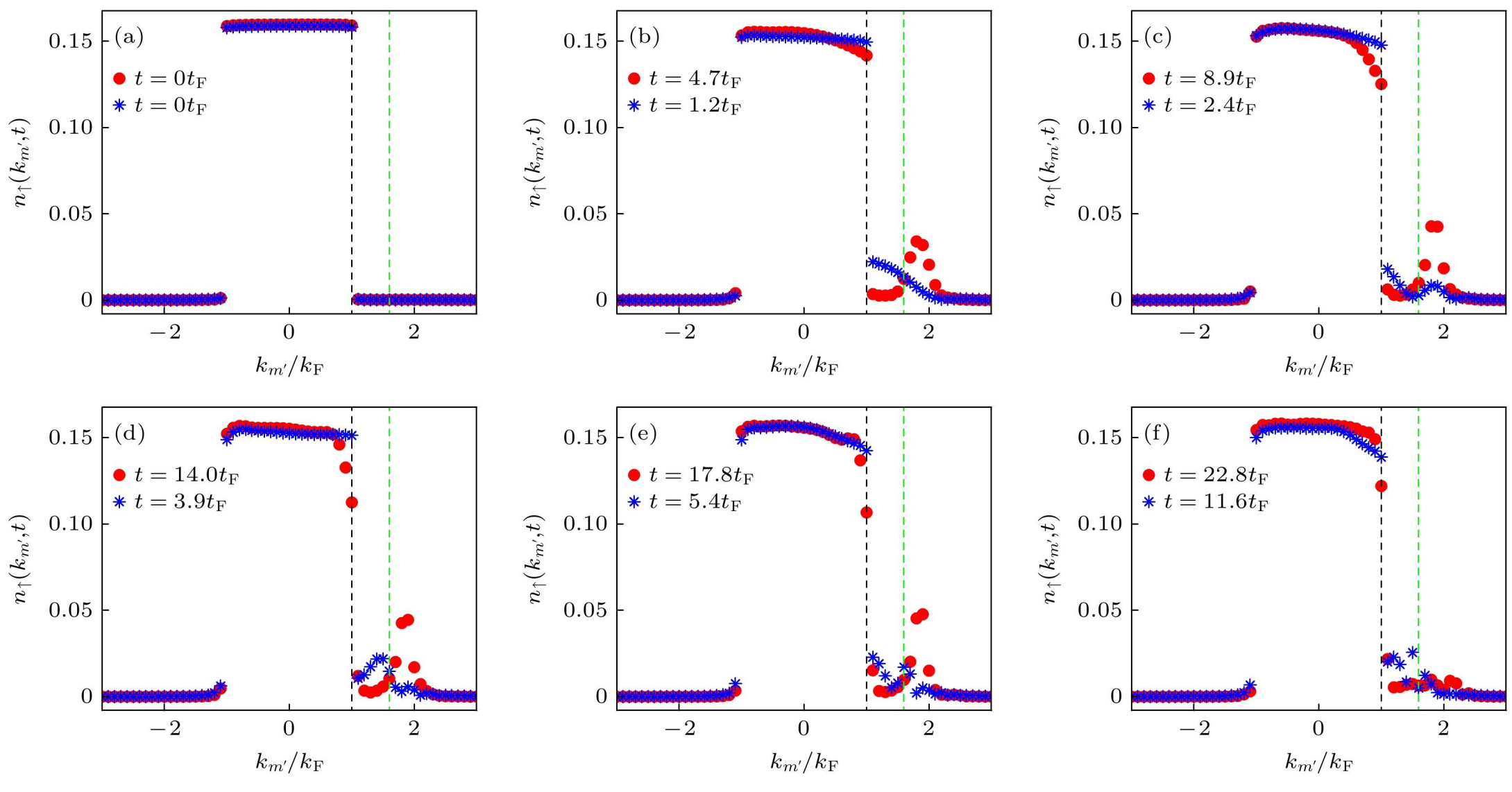

**Fig. 6 Momentum distributions of the spin-up particles at different times for a total momentum $Q$=1.6$k_{\mathrm{F}}$. Red circles correspond to the repulsive interaction case with $c$=5, while blue octagrams correspond to the attractive interaction case with $c$=−5. The green dashed line indicates the position $k_{m'}$=$Q$, and the black dashed line marks the right Fermi momentum. The times chosen in panels (b)–(f) correspond to the coordinates labeled b–f in Fig. 5. The particle number is $N$=22 and the system length is $L$=22.**

Under attractive interactions, the blue octagram symbols in Figure 6 indicate that the momentum evolution of spin-down particles reaches its first minimum at time $t = 1.2t_{\mathrm{F}}$. Due to the attractive interaction, the momentum distribution of spin-up particles differs significantly from that at the initial moment. A distribution decreasing from high to low values appears near the Fermi sea boundary up to $k_{m'} = 2k_{\mathrm{F}}$. Unlike the case of repulsive interactions, no peak emerges in the momentum distribution at this time. At $t = 2.4t_{\mathrm{F}}$, the momentum evolution of spin-down particles reaches the first secondary peak. The momentum distribution of spin-up particles shows a gradual decline from the Fermi sea boundary to $k_{m'} = 1.6k_{\mathrm{F}}$. From $k_{m'} = 1.6k_{\mathrm{F}}$ to $k_{m'} = 2k_{\mathrm{F}}$, the distribution first increases and then decreases, exhibiting a small peak near $k_{m'} = 1.8k_{\mathrm{F}}$. This peak height is significantly lower than that observed under repulsive interactions at $k_{m'} = 1.8k_{\mathrm{F}}$. At $t = 3.9t_{\mathrm{F}}$, the momentum evolution of spin-down particles reaches its second minimum. The momentum distribution of spin-up particles peaks at $k_{m'} = 1.4k_{\mathrm{F}}$, which is slightly lower than the position corresponding to $Q = 1.6k_{\mathrm{F}}$. At $t = 5.4t_{\mathrm{F}}$, the momentum evolution of spin-down particles reaches the second secondary peak, while the momentum distribution of spin-up particles peaks at $k_{m'} = 1.6k_F$. In summary, the peak position in the system is slightly smaller than $k_{m'} = 1.8k_F$ observed in the repulsive case, and the peak height is also lower. This occurs because stronger attractive interactions between particles of different spins increase the effective mass of the particles. Since momentum is conserved during evolution, the effective velocity of the particles decreases. At $t = 11.6t_F$, the momentum distribution of spin-up particles remains high outside the Fermi sea, and the system does not return to a state similar to the initial one. Consequently, the momentum of spin-down particles exhibits only weak quantum revival phenomena.

## 5.2 Quantum Revival Period and Chemical Potential of Spin-Down Particles

As shown in Figure 7(a), the period $\tau$ of quantum revivals is consistent with the period corresponding to the chemical potential $\mu_{\downarrow c}$ of spin-down particles in the ground state with zero total momentum. This behavior is similar to that of repulsive Fermi gases[40], namely

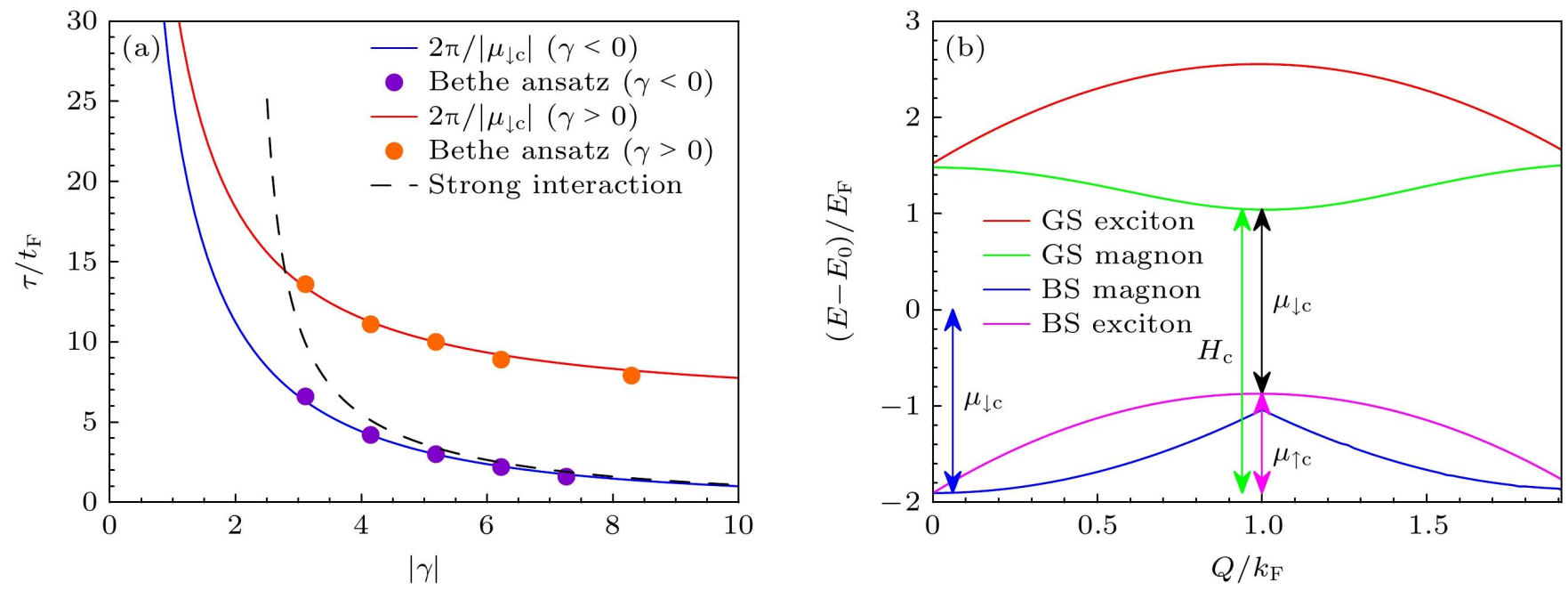

**Fig. 7 (a) Dependence of the quantum flutter period on the dimensionless interaction strength; (b) dispersion relations of the excitonic and magnonic states in the gas-state regime and in the bound-state regime. The dimensionless interaction strength is $\gamma=c/n=-5$. The particle numbers in (a) and (b) are $N=28$ and $N=100$, respectively.**

$$\tau = \frac{2\pi}{|\mu_{\downarrow c}|}, \quad (37)$$

The absolute value of $\mu_{\downarrow c}$ is taken because the chemical potential of spin-down particles is negative under attractive interactions. Various derivation methods yield consistent results for the chemical potential of spin-down particles. At zero total momentum, one can derive the critical magnetic field and critical chemical potential at the quantum phase transition point using thermodynamic Bethe ansatz (BA) equations. Alternatively, discrete BA equations allow derivation based on the energy required to add a spin-down particle to the ground state of an ideal Fermi gas. When the total momentum equals the Fermi momentum, the energy difference between the gaseous magnon state and the bound exciton state equals this chemical potential. When the total momentum exceeds the Fermi momentum, the energy difference between the bound exciton state (which has a high occupation probability and forms pairwise) and the gaseous magnon state equals this chemical potential. This derivation utilizes the relationship between single-particle excitation energy and dressing energy.

We first introduce the method for deriving the chemical potential of spin-down particles using thermodynamic BA equations. The value of $\mu_{\downarrow c}$ is determined by the critical chemical potential $\mu_c$ and the critical magnetic field $H_c$ at the quantum phase transition point:

$$\mu_{\downarrow c} = \mu_c - \frac{H_c}{2}. \tag{38}$$

Meanwhile, the chemical potential $\mu_{\uparrow c}$ of spin-up particles can be expressed in the form of an ideal Fermi gas:

$$\mu_{\uparrow c} = \mu_c + \frac{H_c}{2} = \pi^2 n^2, \tag{39}$$

Here, $n = N/L$. In the thermodynamic limit, where $N \to \infty$ and $L \to \infty$, the critical chemical potential $\mu_c$ and critical magnetic field $H_c$ at the quantum phase transition point can be obtained using the BA integral equations that describe thermodynamic equilibrium states[41].

$$\epsilon^{\mathrm{b}}(\lambda) = 2\left(\lambda^2 - \mu - \frac{c^2}{4}\right) - \int_{-B}^{B} a_2(\lambda - \lambda')\epsilon^{\mathrm{b}}(\lambda')\mathrm{d}\lambda' - \int_{-Q_c}^{Q_c} a_1(\lambda - k)\epsilon^{\mathrm{u}}(k)\mathrm{d}k, \tag{40}$$

$$\epsilon^{\mathrm{u}}(k) = \left(k^2 - \mu - \frac{H}{2}\right) - \int_{-B}^{B} a_1(k - \lambda)\epsilon^{\mathrm{b}}(\lambda)\mathrm{d}\lambda, \tag{41}$$

where

$$a_{m'}(\xi) = \frac{1}{2\pi}\frac{m'|c|}{(m'c/2)^2 + \xi^2}. \tag{42}$$

The dressed states with $\epsilon^{\mathrm{b}}(\lambda) \leqslant 0$ [$\epsilon^{\mathrm{u}}(k) \leqslant 0$] correspond to occupied states where $|\lambda| \leqslant B$ ($|k| \leqslant Q_c$). Positive values of $\epsilon^{\mathrm{b}}$ ($\epsilon^{\mathrm{u}}$) correspond to unoccupied states. The integration boundaries $B$ and $Q_c$ characterize the Fermi surfaces of bound pairs and unpaired fermions, respectively. The transition point from a fully polarized Fermi gas to the emergence of spin-down particles is determined by $B = 0$, $\epsilon^{b}(0) = 0$, and $\epsilon^{u}(\pm Q_{c}) = 0$. This yields the critical magnetic field, specifically

$$H_{c} = 2Q_{c}^{2} + \frac{|c|^{2}}{2} + |c|n - \left(\frac{2Q_{c}^{2}}{\pi} + \frac{|c|^{2}}{2\pi}\right)\arctan\left(\frac{2Q_{c}}{|c|}\right), \tag{43}$$

Here, $Q_{c} = \pi n$. Equation (39) yields $\mu_{c}$. Substituting the critical magnetic field and critical chemical potential into Equation (38) provides the chemical potential for spin-down particles. In the strong interaction limit, the period corresponding to the chemical potential of spin-down particles can be expressed using Equation (37) as

$$\tau = \left|\frac{2\pi t_{F}}{-\frac{\gamma^{2}}{2\pi^{2}} - 1 - \frac{8}{3\gamma} + o(\gamma^{-3})}\right|. \tag{44}$$

Figure 7(a) indicates that the period of quantum flutter approaches zero as $|\gamma| \to \infty$.

The chemical potential of spin-down particles can be understood as the energy required to add one spin-down particle to an ideal Fermi gas system. This corresponds to the value indicated by the blue double-headed arrow in Figure 7(b). Based on this understanding, we obtain:

$$\mu_{\downarrow c} = E_{\mathrm{Ground}}(0) - E_{0}, \tag{45}$$

Here, $E_{0} = \sum_{j=1}^{N-1}(k_{j})^{2}$ and $k_{j} = [-(N/2-1), \cdots, (N/2-1)]2\pi/L$ represent the ground-state energy of an ideal Fermi gas system composed of $N-1$ fermions. For the discrete BA equations (9)-(12) describing bound states, the quantum numbers corresponding to the ground state are $n_{\lambda'} = 0$ and $n_{j'} = -(N-3)/2, \cdots, (N-3)/2$, with the real part of the string solution being $p = 0$. In the thermodynamic limit, the ground-state energy for zero total momentum $(Q = 0)$ is obtained as follows:

$$\begin{aligned} E_{\mathrm{Ground}}(0) = \quad & E_{\mathrm{bare}}(N-2) + \left(\frac{2Q_{c}^{2}}{\pi} + \frac{|c|^{2}}{2\pi}\right)\arctan\frac{2Q_{c}}{|c|} \\ & -|c|n - c^{2}/2 + o(1/L), \end{aligned} \tag{46}$$

Here, the bare energy is given by $E_{\mathrm{bare}}(N-2) = \sum_{j=1}^{N-2}(k_{j'}^{\mathrm{bare}})^{2}$, where the bare momentum is $k_{j'}^{\mathrm{bare}} = 2\pi n_{j'}/L$. Substituting the results obtained by converting the sums of $E_{\mathrm{Ground}}(0)$ and $E_{0}$ into integrals into Eq. (45) yields the chemical potential for spin-down particles as

$$\mu_{\downarrow c} = \left(\frac{2Q_c^2}{\pi} + \frac{|c|^2}{2\pi}\right)\arctan\frac{2Q_c}{|c|} - \left(\pi^2 n^2 + |c|n + \frac{|c|^2}{2}\right). \tag{47}$$

This result aligns with that derived from the BA integral equations describing thermodynamic equilibrium states.

We find that the period of quantum trembling is determined by two states at the Fermi momentum: the gaseous magnon state and the bound exciton state. The energy difference between these two states equals the chemical potential of spin-down particles, corresponding to the value indicated by the black double-headed arrow in Figure 7(b):

$$\mu_{\downarrow c} = E_{\mathrm{BSExciton}}(k_F) - E_{\mathrm{GSMagnon}}(k_F). \tag{48}$$

The magnon state represents a spin-down particle excited from $k = 0$ to other quasiparticle states within the Fermi sea. The exciton state represents a spin-down particle located near $k = 0$, where only one spin-up particle is excited from within the Fermi sea to the Fermi surface, thereby leaving a hole in the Fermi sea. For the ground state of the gaseous magnon state, where the total momentum equals the Fermi momentum ($Q = k_F$), the spin-down particle resides at the Fermi surface. This corresponds to the quantum numbers $n_\lambda = N/2 - 1$ and $n_j = -(N-1)/2, \cdots, (N-1)/2$ in Eqs. (7) and (8) of the discrete Bethe Ansatz equations. By applying the approximation in the thermodynamic limit, $\lambda \to -\infty$, and neglecting higher-order small terms of $1/L$, we obtain the energy eigenvalue:

$$E_{\mathrm{GSMagnon}}(k_F) = E_{\mathrm{bare}}(N) + o(1/L), \tag{49}$$

Here, the bare energy is given by $E_{\mathrm{bare}}(N) = \sum_{j=1}^{N}(k_j^{\mathrm{bare}})^2$, where the bare momentum is $k_j^{\mathrm{bare}} = 2\pi n_j/L$. For the exciton state in a bound state with total momentum equal to the Fermi momentum ($Q = k_F$), spin-up particles are excited from $k = 0$ to the Fermi surface. The quantum numbers in Eqs. (9)—(12) of the corresponding discrete Bethe ansatz equations are taken as $n_{\lambda'} = N/2 - 1$ and $n_{j'} = -(N-3)/2, \cdots, -1/2, 3/2, \cdots, (N-1)/2$. In the thermodynamic limit, $\lambda \to 0$, and the eigenenergy is

$$E_{\mathrm{BSExciton}}(k_{\mathrm{F}}) = E_{\mathrm{bare}}(N) - Q_{\mathrm{c}}^2 + \left(\frac{2Q_{\mathrm{c}}^2}{\pi} + \frac{|c|^2}{2\pi}\right) \times \arctan\frac{2Q_{\mathrm{c}}}{|c|} - |c|n - \frac{|c|^2}{2} + o(1/L). \tag{50}$$

Substituting Eqs. (49) and (50) into Eq. (48) yields $\mu_{\downarrow\mathrm{c}}$. This result is consistent with Eq. (47), which defines the chemical potential for spin-down particles when the total momentum is zero. Compared with the repulsive case[42], both scenarios involve the joint participation of exciton and magnon states. However, they differ in that the repulsive case involves only the gaseous state, whereas the attractive case involves the joint participation of both gaseous and bound states.

Besides the aforementioned method, an indirect approach can determine the energy corresponding to the quantum tremor period when the total momentum equals the Fermi momentum. This method involves calculating the energy differences of the gaseous magnon state and the bound exciton state relative to the ground state, respectively. The target energy is then obtained by taking the difference between these two values. At the Fermi momentum, the energy difference between the bound exciton state and the ground state is

$$\delta E_{\mathrm{BSExciton}}(k_{\mathrm{F}}) = E_{\mathrm{BSExciton}}(k_{\mathrm{F}}) - E_{\mathrm{Ground}}(0) = \mu_{\uparrow\mathrm{c}}, \tag{51}$$

This corresponds to the value indicated by the magenta double-headed arrow in Fig. 7(b). At the Fermi momentum, the energy difference between the gaseous magnon state and the ground state is

$$\delta E_{\mathrm{GSMagnon}}(k_{\mathrm{F}}) = E_{\mathrm{GSMagnon}}(k_{\mathrm{F}}) - E_{\mathrm{Ground}}(0) = H_{\mathrm{c}}, \tag{52}$$

This corresponds to the value indicated by the green double-headed arrow in Fig. 7(b). By subtracting the aforementioned two energy differences and applying Eqs. (38) and (39), we obtain the chemical potential for spin-down particles:

$$\delta E_{\mathrm{BSExciton}}(k_{\mathrm{F}}) - \delta E_{\mathrm{GSMagnon}}(k_{\mathrm{F}}) = \mu_{\uparrow\mathrm{c}} - H_{\mathrm{c}} = \mu_{\downarrow\mathrm{c}}, \tag{53}$$

The results are consistent with Eq. (47).

For momenta exceeding the Fermi momentum, the chemical potential of spin-down particles can still be expressed as the energy difference between the bound exciton state, which features high occupation probability and mutual pairing, and the gaseous magnon state. This formulation further yields the period of quantum oscillations. The excitation energy of the bound exciton state can be understood as the sum of single-particle excitation energies[43], namely

$$\Delta E_{\mathrm{BSExciton}} = \varepsilon^{\mathrm{b}}(\lambda_{\mathrm{b}}) + \varepsilon^{\mathrm{u}}(k_{\mathrm{p}}^{\mathrm{e}}) - \varepsilon^{\mathrm{u}}(k_{\mathrm{h}}^{\mathrm{e}}), \quad (54)$$

Here, $\lambda_{\mathrm{b}}$ denotes the rapidity of the bound pair, $k_{\mathrm{p}}^{\mathrm{e}}$ represents the wave vector corresponding to the particle excited above the Fermi sea, and $k_{\mathrm{h}}^{\mathrm{e}}$ denotes the wave vector corresponding to the hole. The terms $\varepsilon^{\mathrm{u}}(k_{\mathrm{p}}^{\mathrm{e}})$ and $-\varepsilon^{\mathrm{u}}(k_{\mathrm{h}}^{\mathrm{e}})$ represent the excitation energies of the unpaired particle and hole, respectively, while $\varepsilon^{\mathrm{b}}(\lambda_{\mathrm{b}})$ denotes the excitation energy of the bound pair. Since the system contains only one spin-down particle, we neglect the integral term associated with the excitation energy $\varepsilon^{\mathrm{b}}(\lambda_{\mathrm{b}})$ of the spin-down particle. Consequently, the excitation energy of the bound pair in Eq. (40) is expressed as

$$\varepsilon^{\mathrm{b}}(\lambda_{\mathrm{b}}) = 2\left(\lambda_{\mathrm{b}}^2 - \mu - \frac{c^2}{4}\right) - \int_{-Q_{\mathrm{c}}}^{Q_{\mathrm{c}}} a_1(\lambda_{\mathrm{b}} - k)\varepsilon^{\mathrm{u}}(k)\mathrm{d}k, \quad (55)$$

The excitation energy of unpaired particles (background particles), as given in Eq. (41), can be specifically expressed as

$$\varepsilon^{\mathrm{u}}(k) = k^2 - \mu_{\mathrm{c}} - \frac{H_{\mathrm{c}}}{2}. \quad (56)$$

The excitation energy of the gaseous magnon state can also be understood as the sum of single-particle state excitation energies, namely

$$\Delta E_{\mathrm{GSMagnon}} = \varepsilon^{\mathrm{c}}(k_{\mathrm{p}}^{\mathrm{m}}) + \varepsilon^{\mathrm{s}}(\lambda_{\mathrm{s}}). \quad (57)$$

Here, $k_{\mathrm{p}}^{\mathrm{m}}$ denotes the wave vector corresponding to gaseous particles excited above the Fermi sea. The specific expression for the energy $\varepsilon^{\mathrm{c}}(k_{\mathrm{p}}^{\mathrm{m}})$ is identical to Eq. (56) for the bound state. $\lambda_{\mathrm{s}}$ represents the rapidity of spin-down particles, and $\varepsilon^{\mathrm{s}}(\lambda_{\mathrm{s}})$ denotes the spin excitation energy of the gaseous state:

$$\varepsilon^{\mathrm{s}}(\lambda_{\mathrm{s}}) = H_{\mathrm{c}} + \int_{-Q_{\mathrm{c}}}^{Q_{\mathrm{c}}} a_1(\lambda_{\mathrm{s}} - k)\varepsilon^{\mathrm{c}}(k)\mathrm{d}k. \quad (58)$$

Therefore, the energy corresponding to the quantum flutter period can be expressed as the energy difference between the paired bound exciton state and the gaseous magnon state.

$$\Delta E_{\mathrm{QF}} = \Delta E_{\mathrm{BSExciton}} - \Delta E_{\mathrm{GSMagnon}}. \quad (59)$$

In the thermodynamic limit, $\lambda_{\mathrm{s}} \to \infty$, $\lambda_{\mathrm{b}} = 0$, $k_{\mathrm{p}}^{\mathrm{m}} = k_{\mathrm{p}}^{\mathrm{e}}$, and $k_{\mathrm{h}}^{\mathrm{e}} = 0$. Substituting the relevant variables into Eqs. (54) and (57) yields

$$\Delta E_{\mathrm{QF}} = |\varepsilon^{\mathrm{u}}(0)| - |\varepsilon^{\mathrm{s}}(\infty)| = \mu_{\uparrow\mathrm{c}} - H_{\mathrm{c}} = \mu_{\downarrow\mathrm{c}}. \quad (60)$$

This resembles the expression for the chemical potential of spin-down particles in a repulsive Bose gas described in Ref. [7]. Both represent the energy difference between the hole excitation energy of the exciton state and the spin excitation energy of the magnon state. Consequently, the chemical potential of spin-down particles in this scenario simplifies to a form consistent with Eq. (38).

# 6 Conclusion

Based on the exact Bethe ansatz solution for a one-dimensional Fermi gas with attractive interactions and a single spin flip, we reduced the overlap integrals and the matrix elements of single- and two-body correlation functions to finite sums of simple functions. This approach enables efficient calculation of occupation probabilities for numerous eigenstates and the long-time evolution of correlation functions. We injected a spin-down impurity particle with momentum into the ground state composed of non-interacting spin-up fermions. Subsequently, we quenched the interaction between particles of different spins to a finite attractive strength at the initial time. We systematically investigated the evolution characteristics of two-body correlation functions and quantum quenching behavior under varying interaction strengths. First, under weak attractive interactions: when the initial momentum is less than or equal to the Fermi momentum, the system dynamics are dominated by bound states. The impurity particle forms bound pairs with background particles. The two-body correlation function exhibits significant probability density peaks in the local correlation region. Furthermore, correlation holes moving in the positive direction form near these peaks. The evolution period of these holes corresponds to the period associated with the energy difference between the two bound states with the highest occupation probabilities. When the initial momentum exceeds the Fermi momentum, the coupling among unconventional states, gas-like states, and bound states causes the correlation function to alternate between peaks and troughs over time in the local correlation region. Values greater than 1 at the peaks indicate bound-state characteristics, reflecting that impurity and background particles form bound pairs and move cooperatively. Values less than 1 at the troughs indicate gas-like characteristics, reflecting the dissociation of bound pairs during motion. Due to the joint participation of multiple eigenstates with high occupation probabilities, the evolution of correlation peaks cannot be described solely by the two states with the highest occupation probabilities. Second, under strong attractive interactions: regardless of whether the initial momentum of the impurity particle exceeds the Fermi momentum, the peak value of the correlation function in the local correlation region remains greater than 1, exhibiting bound-state characteristics. Additionally, correlation holes that do not propagate outward form on both sides of the local correlation peak. Finally, we investigated the evolution of the momentum of spin-down particles under attractive interactions. We further established a unified relationship between the quantum quenching period and the chemical potential of spin-down particles for arbitrary interaction strengths, namely $\tau = 2\pi/|\mu_{\downarrow c}|$. We obtained consistent values for the chemical potential of spin-down particles using various methods, including thermodynamics and discrete Bethe ansatz equations. This study provides dynamical evolution results in the fields of quench dynamics with attractive interactions and quantum quenching. These findings help deepen the understanding of non-equilibrium properties in quantum many-body physics and offer theoretical references for related experiments on impurity dynamics in ultracold atoms.